\PassOptionsToPackage{dvipsnames}{xcolor}
\documentclass[conference]{IEEEtran}

\usepackage{booktabs} 

\usepackage{subfiles}
\usepackage{graphicx}
\usepackage{array}
\usepackage{textcomp}
\usepackage{listings}
\usepackage{parcolumns}
\usepackage{subfig}
\usepackage{tabularx}
\usepackage{array}
\usepackage{url}
\usepackage{xspace}
\usepackage{silence}
\usepackage[sort, numbers]{natbib}

\lstdefinestyle{motivatingExample}{
	basicstyle=\footnotesize\tt,
	language = Java,
	numbers = left,
	stepnumber = 1,
	numbersep = 10pt,
	tabsize = 2,
	breaklines= true,
	showspaces = false,
	showstringspaces = false,
	keywordstyle=\color{blue}\bfseries,
	columns=fullflexible,
    frame=b,
    linewidth=\textwidth,
    framexleftmargin=0.35in
}

\lstdefinestyle{regular}{
	basicstyle=\footnotesize\tt,
	language = Java,
	numbers = left,
	stepnumber = 1,
	numbersep = 10pt,
	tabsize = 2,
	breaklines= true,
	showspaces = false,
	showstringspaces = false,
	keywordstyle=\color{blue}\bfseries,
    xleftmargin=5.0ex,
    framexleftmargin=0.35in
}

\newcolumntype{L}[1]{>{\raggedright\arraybackslash}p{#1}}
\newcolumntype{R}[1]{>{\raggedleft\arraybackslash}p{#1}}
\newcolumntype{C}[1]{>{\centering\arraybackslash}p{#1}}

\usepackage{xargs}
\usepackage[colorinlistoftodos,prependcaption,textsize=small,textwidth=0.64in]{todonotes}
\newcommandx{\unsure}[2][1=]{\todo[linecolor=red,backgroundcolor=red!15,bordercolor=red,#1]{#2}}
\newcommandx{\mytodo}[2][1=]{\todo[linecolor=orange,backgroundcolor=orange!25,bordercolor=orange,#1]{#2}}
\newcommandx{\later}[2][1=]{\todo[linecolor=blue,backgroundcolor=blue!15,bordercolor=blue,#1]{#2}}
\newcommandx{\info}[2][1=]{\todo[linecolor=Plum,backgroundcolor=Plum!10,bordercolor=Plum,#1]{#2}}
\newcommandx{\edit}[2][1=]{\todo[linecolor=OliveGreen,backgroundcolor=OliveGreen!10,bordercolor=OliveGreen,size=\normalsize,#1]{#2}}
\newcommandx{\thiswillnotshow}[2][1=]{\todo[disable,#1]{#2}}

\newcommand{\dpspec}{\textsc{DPSpec}\xspace}

\newcommand{\droidperm}{\textsc{DPerm}\xspace}
\newcommand{\MyPara}[1]{\textbf{#1:}}

\begin{document}

\title{\droidperm: Assisting the Migration of Android Apps to Runtime Permissions}

\author{\IEEEauthorblockN{Denis Bogd\u{a}na\c{s}}
\IEEEauthorblockA{Oregon State University\\
bogdanad@oregonstate.edu}
}
 
\maketitle
\thispagestyle{plain}
\pagestyle{plain}

\begin{abstract}
Android apps require permissions when accessing resources related to privacy or system integrity.
Starting from Android 6, these permissions have to be asked at runtime.
However, migrating to the new permission model poses multiple challenges for developers.
First, developers have to discover where the app uses permissions, which requires a permission specification.
To date several such specifications have been built,
yet these are either imprecise, incomplete or don't support all types of protected resources.

We first present \dpspec, a novel permission specification built from several documentation formats supplied with the Android SDK.
Compared with the state the art specification, it contains 2.5x as many entries for protected methods and detects dangerous permission usages in more than twice as many apps.

A second challenge for developers is where to insert permission requests, with possible locations restricted by the request mechanism.

We also present \droidperm\footnote{Renamed to comply with submission policy.}, a static analysis for Android apps that recommends locations for permission requests in code.
It achieves high precision through context sensitivity and improves recall through a general call graph augmentation algorithm for incomplete code.
Our empirical evaluation on 32 apps shows a precision of 96\% and recall of 89\%.

\end{abstract}

\setlength\emergencystretch{\hsize}
\hbadness=10000

\section{Introduction}\label{sect:introduction}


Android apps require permissions to access resources related to user privacy or system integrity, such as Camera, GPS, Calendar.
Up to Android 5, the user was presented with the list of required permissions at app installation time.
He had only two choices: to grant all requested permissions and to install the app,
or to reject them and cancel the installation.

This static permission model had security and privacy issues.
It made it easy for apps to ask for more permissions than needed.
The problem was exacerbated by incomplete permission documentation \cite{perm-msr-13} that made developers unsure of what permissions their app really needs.
As a result, about one third of apps were over-privileged~\cite{perm-demystified}.
This increased the chance that malware can abuse granted permissions to perform malicious activities~\cite{perm-risks-benefits, perm-sleeping-android}.
Even when apps used granted permissions, users might have disagreed with the way they are used.
A study conducted on Android 4.1~\cite{perm-remystified} shows that users would block one third of app accesses to protected resources, if they would have this choice.


To address these issues, the permission model was overhauled in Android 6.
Apps now ask permissions as needed, at runtime.
Users can now make informed decisions whether to grant permissions based on the context in which permissions are used, reducing the risk of abuse.

However, migrating apps to the new model requires non-trivial effort from app developers.
Next we describe the two main challenges they face, and our proposed solutions.

\subsection{The Permission Specification}


The first challenge for developers is to identify where their app uses protected resources.
These resources represent specific Android SDK methods or fields.
The problem is complicated by the lack of a centralized permission specification for Android~\cite{so-sensitive-list},
consisting of mappings from protected resources to permissions.
In addition, the problem is complicated by incomplete documentation.
Exhaustively testing the app in attempt to detect crashes might not be enough, because most protected resources don't throw exceptions if they are accessed without permissions.
Instead, they return null or empty results.
Overwhelmed by these problems, developers often ask help on StackOverflow~\cite{so-sensitive-list-2,so-sensitive-list-3}.

Building a permission specification for Android is an active area of research.
It started with Stowaway~\cite{perm-demystified}, which inferred required permissions through unit testing and feedback directed API fuzzing.
PScout~\cite{pscout} produced the specification through reachability analysis of Android Framework code, on a simple class hierarchy(CHA)-based call graph.
PScout specification was significantly more complete.
However, evaluation of the most recent state of the art Axplorer~\cite{axplorer} shows that PScout specification suffers from imprecision.
Axplorer adds a number of enhancements on top of PScout that dramatically improve precision.
In addition, Axplorer is the only work in this list that supports Android 6.
Yet, as we show in this paper, Axplorer suffers instead from significant incompleteness.
In addition, it only supports protected methods, but not fields.

For these reasons we decided to build our own permission specification, hereafter called \dpspec.
Unlike the previous work~\cite{perm-demystified,pscout,axplorer}, we collected it from various documentation formats provided by Android: Java annotations, XML annotations and Javadoc comments.
We use a semi-automated approach that involves both automated tools for collecting references to permissions, as well as manual inspection.
When facing ambiguities, we interpreted the documentation conservatively, according to a precise methodology.
In addition, we inferred a small number of mappings by analyzing the source code of 22 apps from \url{f-droid.org}, an app store alternative to Google Play.
In this process we built a highly precise and most complete specification for dangerous permissions currently available, and the first to include protected fields.

We also compared \dpspec and Axplorer specification in terms of absolute size and coverage of protected resource usages in apps.
In size comparison, \dpspec contains 120 mappings for protected methods alone, compared to 45 for Axplorer.
In coverage comparison, \dpspec helps detecting protected method usages in more than twice as many apps as Axplorer.
Both comparisons provide evidence that \dpspec is more complete.

\subsection{Locations for Permission Requests}
The second challenge faced by developers is where to insert permission requests.
Due to their mechanism of action, permission requests can typically be placed only at the end of callbacks --- events initiated by the system.
Thus, developers have to propagate the requests from places where the protected access is necessary to places where a request can be performed.
This can be non-trivial when protected resources are nested deep inside call chains that extend into third party code.
Possibly for these reasons, as of Nov. 2016, only 25\% of the 1980 apps from f-droid.org were migrated to Android 6, as we show in this paper.


We present \droidperm, a static analysis tool for recommending locations where permission requests have to be inserted.
To achieve high precision, it performs reachability analysis starting from callbacks, on a context sensitive call graph.
A special challenge when analyzing Android apps is lack of access to Android SDK source code, which could lead to vastly incomplete call graphs and consequently low recall.
To mitigate this difficulty, \droidperm augments the call graph with safe edges where possible, using a simple re-usable algorithm.

We rigorously evaluated precision and recall of \droidperm on a corpus of 32 apps, using a combination of specialized static analysis tools and manual inspection.
\droidperm achieves a precision of 96\% and recall of 89\%.
We also provide justification for instances when false positives or negatives were found.
Moreover, we evaluated the effect of context-sensitivity on the tool's precision and
we found that context-sensitivity reduces the number of false positives 4-fold.

\subsection{Contributions}

This paper makes the following main contributions:
\begin{enumerate}
    \item \textbf{\dpspec:} the most complete specification for dangerous permissions currently available (\S\ref{sec:perm_mapping}).
    \item \textbf{\droidperm:} A static analysis tool for recommending permission request locations (\S\ref{sec:sensitives-analysis}).
    \item \textbf{Empirical evaluation:} We rigorously measure the precision and recall of \droidperm on a corpus of real-world apps (\S\ref{sec:eval-analysis}).
\end{enumerate}

and the following side contributions:
\begin{enumerate}
  \item A comparison in size and coverage between \dpspec and Axplorer specification (\S\ref{sec:axplorer}).
  \item A small scale study of declaratively over-privileged apps (\S\ref{sec:over-privileged-apps}).
\end{enumerate}

We plan to publish our tools, the evaluation corpora, and the detailed empirical results\footnote{Removed to comply with double blind submission policy.}.

\section{Background}\label{sec:background}

\begin{figure}
\begin{lstlisting}[style=regular]
import static android.Manifest.permission.CAMERA;
import static android.content.pm.PackageManager.PERMISSION_GRANTED;

public class CameraActivity extends Activity {
  private static final int REQ_CODE = 1;

  protected void onCreate(...) {
    if (checkSelfPermission(CAMERA)
        != PERMISSION_GRANTED) {
      requestPermissions(
          new String[]{CAMERA}, REQ_CODE);
    } else {
      takePicture();
    }
  }

  private void takePicture() {
    if (USE_INTERNAL_CAMERA) {
      CameraManager cm =
        this.getSystemService(CameraManager.class);
      cm.openCamera(...); // method sensitive
      ...
    } else {
      startActivity(
        new Intent( //parametric sensitive
          MediaStore.ACTION_IMAGE_CAPTURE//field sensitive
        ));
      ...
    }
  }

  public void onRequestPermissionsResult(int reqCode, ..., int[] grantResults) {
    if(reqCode == REQ_CODE
        && grantResults[0]== PERMISSION_GRANTED) {
      takePicture();
    } else {
      //show explanation msg and close activity
    }
  }
}
\end{lstlisting}
\caption{Permission request example with method and field sensitives}
\label{f:background_example}
\end{figure}

Up to Android 5, apps declared permissions in their manifest file,
and were granted permissions during installation.
This often led to well known privacy and over-privilege issues~\cite{perm-demystified}.
To give users more control, starting with Android 6, permissions were divided into two categories.
Normal permissions, like \verb|VIBRATE| or \verb|INTERNET| are still granted at installation.
Dangerous permissions~\cite{url-normal-and-dangerous-perm}, which control access to privacy-sensitive resources, have to be requested at runtime, ideally immediately before accessing a feature that requires them~\cite{url-perm-best-practices}.
There are 24 dangerous permissions in total; examples are \verb|CAMERA|, \verb|ACCESS_FINE_LOCATION|, \verb|READ_CALENDAR|.

A typical permission request pattern is shown in Figure~\ref{f:background_example}.
Method \verb|onCreate()| is a callback, e.g. is system-invoked.
On line 8 the app checks whether it has access to \verb|CAMERA|.
If it doesn't, it requests the permission through \verb|requestPermissions()| and the current callback ends.
Method \verb|requestPermissions()| does not request permissions immediately.
Instead, it schedules a permission request event, to be executed immediately after current callback terminates.
For this reason, any code that follows after permission request cannot depend on these permissions.
Most typically, permission request is the last code to be executed in the callback.
If on line 8 permission is already granted, \verb|takePicture()| is invoked, where camera access happens.

After the permission request event, the system calls \verb|onRequestPermissionsResult()| on line 32.
If permissions are granted, apps execute the code that uses them. In our case it is the same \verb|takePicture()|.
If permissions are rejected, apps typically display an explanation message and revert to state prior to the action that initiated permission request.
Alternatively, apps might disable the functionality requiring permissions, or shut down entirely, if permissions are critical for basic functioning.

In this paper we call code locations that access protected resources \textit{sensitives}.
Method \verb|takePicture()| contains two kinds of sensitives: a \textit{method sensitive} on line 21 and an \textit{field sensitive} on line 26.
As expected, method sensitives are associated with specific methods, while field sensitives - with specific fields.
In addition, we call \textit{parametric sensitives} the set of methods that can take sensitive fields as arguments.
In the example, the parametric sensitive is the constructor of \verb|Intent|.
The code requiring permissions is actually the call to parametric sensitive, if the parameter is a protected field.

There are several types of field sensitives, based on their broad functionality. The first type is intent actions.
They are used to start other activities.
In our example, lines 24-27 open a system-managed activity to take a picture.
The second type is associated with broadcast receivers and are used to receive external notifications.
And the third type is URI sensitives, which identify various collections of data provided by the system, such as contacts or calendar.
In addition, we identified a few specialized cases where a method requires permissions depending on received arguments.
We modeled them similarly.

Method and field sensitives map to 22 out of 24 dangerous permissions documented in~\cite{url-normal-and-dangerous-perm}.
They can be detected through customized reachability analysis and are the focus of the present paper.
The remaining two are storage permissions, not associated with a particular Android SDK field or method.
They are required when app accesses a file outside internal cache directory.
Precisely detecting storage sensitives requires distinguishing between internal and external file accesses.
This in turn requires full-fledged dataflow analysis; we leave it for future work.

As a last note, the behavior when a sensitive is accessed without permissions is very inconsistent.
This is an extra challenge for migrating apps.
Only a small subset of sensitives throw \verb|SecurityException|, making them easy to spot and fix.
Other sensitives return either \verb|null| or an empty list.
And for many broadcast receivers, if permissions are not granted, the app simply doesn't receive any notifications.
These design decisions were made to allow revoking permissions for Android 5 apps when run on Android 6 phones.
Yet on long term this only complicated proper testing and increased the likelihood that apps might end up improperly migrated.

\section{Mining the Permission Specification}\label{sec:perm_mapping}


In this section we present \dpspec, a permission specification collected from multiple annotation and documentation sources in Android SDK.

\subsection{Mappings from Android Annotations}
\label{sec:spec-anno}
With Android 6, Google recognized the need to formally document permissions.
They introduced two formats for this.
One is through Java annotation \texttt{@requirePermissions} attached to a protected method or field.
It can be used in either Android SDK, Google libraries or 3rd party libraries.
The second format is through a set of XML files supplied with Android SDK, which essentially duplicate Java annotations in the SDK, for technical reasons.
The two versions are not necessarily in sync, the XML version being more often updated.\footnote{Discussions with Google engineers, 2017.}
To get the most complete and up to date specification, we decided to use XML metadata for Android SDK and Java annotations for Google Play Services library.
After filtering out non-dangerous permissions, we produced 44 mappings for Android SDK and 46 for Google Play.

\subsection{Coverage Evaluation}
\label{sec:coverage-evaluation}
Unsure of how complete is this specification, we decided to evaluate at high level how many sensitives used in apps have correspondents in \dpspec.
For this we used the following key observation.
In the most common usage pattern, a dangerous permission should appear in 3 domains (places) inside an app.
First, it should be declared in the manifest file:

\begin{lstlisting}[style=regular,numbers=none]
  <uses-permission android:name=
    "android.permission.ACCESS_FINE_LOCATION"/>
\end{lstlisting}

Second, it should appear inside a permission request:

\begin{lstlisting}[style=regular,numbers=none]
  activity.requestPermissions( new String[]{
    Manifest.permission.ACCESS_FINE_LOCATION}, ...);
\end{lstlisting}

Third, the app should have a sensitive that consumes this permission, like:

\begin{lstlisting}[style=regular,numbers=none]
  locationManager.getLastKnownLocation(...)
\end{lstlisting}

If a permission appears in the manifest and the request, but there are no sensitives to consume it, this is a strong indication that the app indeed contains a sensitive for this permission, but this sensitive is not in \dpspec.

\MyPara{Permission Collector tool}
We implemented a simple tool that analyzes a list of apps (through their apk files) and enumerates the permissions used in the three locations above.
The tool analyzes the code in the app and in any linked 3rd party libraries, but not in Android SDK.
For simplicity, instead of permission requests (place 2) it counts all references to a particular permission in the code.
For the purpose of coverage evaluation, we are interested in permissions having one of the following two usage scenarios.
If, in a certain app, a permission appears in the manifest, in Java code (presumably permission request) and is consumed by a sensitive,
we call it a \textit{MCS permission instance}.
If the permission appears in the manifest and code only, it is a \textit{MC permission instance}.

\MyPara{F-droid apps corpus}
For all the needs in this paper, we extracted a corpus of 106 apps from \url{f-droid.org}, a popular unofficial app store.
For this study, f-droid has a number of advantages over Google Play.
(i) It contains only open-source apps.
(ii) There are no restrictions against downloading all the apps or a subset using a batch tool. Google Play does not allow this.
(iii) For every apk build f-droid provides the corresponding source code.
Thus, after performing some static analysis on the apk file, it is easy to do a follow-up manual inspection of the source code.
At the moment the apps were extracted (Nov 2016) the store contained 1980 apps.
Out of them, 500 apps (25\%) were migrated to Android 6 or later.
Most apps don't use any dangerous permissions.
Our corpus consists of the subset of apps satisfying the following conditions:
(a) they declare target API level 23, e.g. Android 6.
and (b) they have at least one dangerous permission other than storage, declared in the manifest.
F-droid also contains apps with target API level above 23.
We did not include them, to avoid dealing with multiple versions of Android.

\MyPara{First coverage results}
The results of running the permission collector over the corpus were rather unexpected.
There were 80 MC permission instances across 54 apps, and only 44 MCS instances across 33 apps.
We define the permissions coverage as the number of MCS instances in the corpus divided by the number of MC+MCS instances.
Consequently, the coverage was only 35\%.
Our specification was vastly incomplete.
As we later discovered from discussions with Google engineers, annotations only cover a subset of sensitives that throw \verb|SecurityException| when called without permissions.
They don't cover sensitives with other behavior mentioned in \S\ref{sec:background}.

\subsection{Mappings from Javadoc}
\label{sec:spec-javadoc}
After briefly analyzing a few apps it was clear that Android SDK contains many more protected members, with permission requirements documented in Javadoc.
We built another tool, Javadoc Miner, to mine permission mappings from Javadoc semi-automatically.
It analyzes Javadoc comments in Android SDK source code, looking for permission identifiers.
The tool collects a map from Java elements to permissions mentioned in their Javadoc.
Such elements could be either methods, fields or classes.
Some identifiers, like \texttt{ACCESS\_FINE\_LOCATION}, are uniquely associated with permissions.
More generic ones, like \texttt{CAMERA}, could appear in multiple contexts.
At least for this reason, a manual inspection was required to decide which comments really represent permission requirements.
After inspecting all discovered methods and fields with permission
references, we added to \dpspec 47 mappings.

In addition to those, Javadoc Miner produced 11 mappings for classes.
They had to be converted to mappings for member fields/methods to be usable.
Therefore, to infer the members, we manually inspected Javadoc documentation in sensitive classes.
In a few classes, like \texttt{SipManager} sensitive members are explicitly mentioned in Javadoc.
For \texttt{SubscriptionManager}, Javadoc states that all public methods are sensitives.
But for most classes there are no such clarifications.
For them we conservatively collected all members whose functionality indicates access to a protected resource, that is not mandatory preceded by a call to another sensitive.
For example for class \texttt{Camera} we only added mappings for two versions of static method \texttt{open} that return an object of type \texttt{Camera}, but not for instance methods.
In order to access any instance method, a call to \texttt{open} would be required first, thus instance methods were not primary sensitives.
Finally, for classes that primarily contain URI sensitives, like \texttt{ContactsContract}, we added mappings for every public field of type \texttt{URI} they contain.
Any of them can be used independently of other sensitives to access protected data.
In total, we extracted 69 additional mappings for permissions documented at class level.

When running permission collector on the extended specification, there were 27 MC permission entries left across 22 apps.

\subsection{Mappings inferred from apps}
\label{sec:spec-apps}
One more way to improve \dpspec was to inspect the source code of the 22 apps above and to search for sensitive candidates.
For each sensitive candidate we found, we searched it's name on StackOverflow, to confirm our finding.
We looked for code examples where author explicitly states that certain permissions are required to access the sensitive candidate.
If we could find such examples, we would add the candidate to \dpspec.
In some cases, when we discovered a class with multiple URI fields that access the same broad protected resource, we added all of them to \dpspec.
This way we produced 32 more mappings.
Full specification thus far had 120 mappings for methods and 118 for fields.
Running coverage analysis produced 18 MC instances across 13 apps, and 106 MCS entries across 75 apps.
Thus the coverage was 85\%.

These remaining 18 MC instances could be divided into following categories. Either the app really doesn't contain sensitives for the permission, suggesting that it might be under development.
Or the app communicates with other Android services through JavaScript code, and sensitives are presumably located in those services.
In some cases apps directly accesses native code related to protected resources.
And we found one case when the app would access sensitive resources in the Android framework layer, which is not part of public Android SDK, through reflection.
All these cases are outside the scope of the present study.


\section{\droidperm: the Sensitives Analysis}\label{sec:sensitive_analysis}

\label{sec:sensitives-analysis}

The purpose of \droidperm is to help migrating apps from Android 5 to Android 6 by recommending permission request insertion points.
Due to the way permission requests work (\S\ref{sec:background}), the safest place to insert them is directly inside callbacks.
In a nutshell, \droidperm is a reachability analysis built on top of Soot~\cite{soot} and FlowDroid~\cite{flowdroid} ecosystem.
The analysis starts from callbacks --- a set of methods in the app that are called by Android framework when various user- or system-generated events occur.
For every callback, \droidperm performs an inter-procedural call graph traversal and determines the set of reachable sensitives.
These in turn lead to specific lines of code and permissions that have to be checked inside each callback.
For evaluation purposes, (\S\ref{sec:eval-analysis}) \droidperm also outputs call paths from callbacks to sensitives.

At high level our analysis consists of three steps. First, \droidperm generates an entry point for the analyzed app, suitable for call graph generation (Subsection~\ref{sec:entry-point-generation}). It then passes the entry point to out of the box tools, to generate a call graph and a context-sensitive points-to analysis.
Finally, \droidperm performs reachability analysis on the call graph
and infers permission request insertion points (Subsection~\ref{sec:guard-insertion-inference}).
Additional subsections describe various other aspects of our analysis.

\subsection{Entry point generation}
\label{sec:entry-point-generation}
Reachability analysis for Java typically starts from the main method.
But there is no single main method in Android.
Instead, an app contains a large number of callbacks in multiple components, invoked by Android system.
A typical way to bridge the gap is to generate a main method that simulates invocations of all callbacks.

\droidperm uses a custom version main method generation algorithm from FlowDroid.
FlowDroid algorithm strives to accurately model the execution order of various callbacks inside Activity.
This is required for flow-sensitive taint analysis.
Yet our early experiments showed that it misses more than half of the callbacks. This is because FlowDroid main method generation does not support \verb|Fragment| component and many Android callbacks defined in interfaces.

Conversely, for \droidperm we strived to achieve a good recall, while using an Android framework model just good enough for our goal.
For this reason we designed a simpler algorithm that traverses all classes in the app and collects the callback methods, defined as following.

A method \verb|C.f()| is considered callback by \droidperm if it either:
\begin{itemize}
	\item Overrides a method from an Android class.
	\item or:
	\begin{itemize}
		\item Implements a method from an Android interface \verb|I| and
		\item There is an object of type \verb|C| that is passed as argument to a method expecting a formal parameter of type \verb|I|.
	\end{itemize}
\end{itemize}

This might look like an over-approximation, but we found that any time an app-level class overrides a method from Android SDK, this method is expected to be called by the framework.
There is only one case to our knowledge when treating such methods as callbacks is undesirable: AsyncTask and other asynchronous constructs, which we discuss later.

For the purpose of callback detection, Android classes/interfaces are all those located in packages \verb|android.| and \verb|com.google.android|, except aforementioned asynchronous constructs.
The only source of imprecision that we are aware of is including in the analysis unused code, e.g. classes derived from Android SDK classes but not used in the app. We leave this case for future work.

After collecting the list of callbacks, DP-detect passes it to {FlowDroid} infrastructure to generate the main method.

\subsection{Modeling sensitives}
\droidperm models both method and field sensitive instances in a unified way, as call sites.
Method sensitive instances are, expectedly, any calls to methods that require permissions.
Field sensitive instances are calls to parametric sensitives (\S\ref{sec:background}) whose sensitive argument may be a field requiring permissions. To illustrate this with an example, in the code below only the first line is a sensitive:

\vspace{0.5em}
\begin{verbatim}
      new Intent(SENSITIVE_FIELD);
      new Intent(SAFE_FIELD);
      print(SENSITIVE_FIELD);
\end{verbatim}

Here constructor of \verb|Intent| is a parametric sensitive and \verb|SENSITIVE_FIELD| is the field requiring permissions.

When sensitive argument is a local variable, we perform an intra-procedural dataflow analysis to determine possible values assigned to the variable. If any of them is a field with permissions, the analyzed call site is considered a sensitive.

\subsection{Reachability analysis}
\label{sec:guard-insertion-inference}

For precise reachability analysis \droidperm needs a context-sensitive call graph. Unfortunately Soot framework doesn't provide suitable out of the box solutions.
\footnote{In the past Soot included PADDLE\cite{paddle} context-sensitive call graph constructor, but it is incompatible with latest Soot and has poor performance.}
Instead, it includes SPARK~\cite{SPARK}, a widely used context-insensitive call graph constructor, and GEOM~\cite{geom}, a 1-CFA context-sensitive points-to analysis that uses SPARK under the hood.
By combining these two tools we achieved the desired level of precision.

After generating the entry point, \droidperm builds a call graph and points-to analysis.
It then performs a forward reachability traversal from each callback to sensitives.
The tool then compiles collected paths into a report containing the lines of code directly inside each callback, for which permissions are required.

To achieve context sensitivity, the reachability algorithm refines the context-insensitive set of edges produced by SPARK with the context-sensitive points-to data given by GEOM.
More precisely, if a method call \verb|a.f()| in the call graph contains edges to two implementations \verb|A.f()| and \verb|B.f()|, but points-to data for \verb|a| in the current context equals \verb|{B}|, then \droidperm only traverses the edge to \verb|B.f()|.


\begin{figure}
\begin{lstlisting}[style=regular,numbers=none]
void callback1() {
  new Thread(new Runnable() {void run() {
    SENSITIVE();
  }}).start();
}

void callback2() {
  new Thread(new Runnable() {void run() {
    safe();
  }}).start();
}
\end{lstlisting}
\caption{Code example requiring 1-CFA context sensitivity}
\label{f:thread-example}
\end{figure}

Context sensitivity is essential for handling Java and Android asynchronous constructs, such as
\verb|Thread|, \verb|ExecutorService|, \verb|Handler| and \verb|AsyncTask|.
Consider the example in Figure~\ref{f:thread-example}.
Without context sensitivity, \verb|Thread.start()| in both contexts would reach the sensitive method, flagging both callbacks as requiring permisisons.
1-CFA level, combined with customizations presented next, is enough to resolve such ambiguities.


\subsection{Call graph augmentation with safe edges}
\label{sec:cg-augmentation}

The source code supplied with Android SDK for packages \verb|android.| and \verb|java.| does not contain any implementation.
Instead, it contains just the members visible to the user, with stub, e.g. empty implementations of all methods.
This lessens computation effort for call graph generation, but the result is vastly incomplete.
In particular, SPARK will not generate any edges for a method call \verb|a.f()| if no allocation site can be inferred for values of \verb|a|.
At the same time, most method and parametric sensitives are instance methods for classes instantiated within the framework, like \verb|CameraManager| in Figure~\ref{f:background_example}.
With only the tools presented so far, almost no sensitives would be reached.

\begin{figure}
\begin{lstlisting}[style=regular,numbers=none]
class MyActivity extends Activity {
  void onCreate() {
    MyView v = this.findViewById(...);
    v.callSensitive();
  }
}

class MyView extends View {
  void callSensitive() {
    LocationManager lm = this.getContext().getSystemService(...);
    lm.getLastLocation();
  }
}
\end{lstlisting}
\caption{Example sensitive reachable after call graph augmentation}
\label{f:find-view-by-id}
\end{figure}

In addition, many app-defined classes are instantiated by the framework, based on various XML configuration files.
We present an example in Figure~\ref{f:find-view-by-id}, distilled from real-life code.
Object \verb|v| is instantiated by the system inside \verb|findViewById|.
Since \droidperm doesn't have access to the implementation of \verb|findViewById|, SPARK won't be able to find allocation sites for \verb|v|, and won't create the edge for \verb|callSensitive()|.
The sensitive \verb|lm.getLastLocation()| will be unreachable.
This limitation was also acknowledged by other research using call graph reachability~\cite{revDroid}.

To partially alleviate the issue, \droidperm augments the call graph with additional edges for every call site that (a) doesn't have any edges produced by SPARK, and (b) can be unambiguously resolved to only one method based on class hierarchy alone.
This operation is done as post-processing, after SPARK execution.

The call graph still doesn't contain edges for polymorphic call sites which could be properly processed by SPARK, but can only be reached through other edges, added during post-processing.
We expect that this last limitation can be fixed if augmentation technique would be integrated directly into SPARK.
We plan to implement this, yet it is a non-trivial engineering task, outside the scope of this paper.

\subsection{Modeling asynchronous constructs}
\label{sec:classpath-crafting}


To properly analyze asynchronous constructs, we replaced stub implementations with hand-crafted implementations,
containing just enough information to generate expected call graph edges.

For example, we crafted the following  straightforward implementation for \verb|Thread.start()|:

\begin{lstlisting}[style=regular,numbers=none]
void start() {
  target.run();
}
\end{lstlisting}

Note that it makes the caller of \verb|start()| the immediate context of statement \verb|target.run()|. 
As a result, 1-CFA context sensitivity is enough to properly traverse the code in Figure~\ref{f:thread-example}.

\section{\droidperm Evaluation}\label{sec:eval-analysis}

In our evaluation we address the following research questions:

\textbf{RQ1: What is the recall of \droidperm?}

\textbf{RQ2: What is the precision of \droidperm?}

\textbf{RQ3: What is the effect of 1-CFA points-to refinement on precision?}

\MyPara{Corpus}
We picked all apps from our permission mining corpus (\S\ref{sec:coverage-evaluation}) that only contain MCS permission instances.
These are the apps most likely to be properly migrated to Android 6 and use only the standard permission request pattern, that we support.
This way we avoided improperly migrated apps, apps that contain many unreachable sensitives, apps that use permissions with inter-app communication or the apps for which we could not find all the sensitives (\S\ref{sec:spec-anno}).
This resulted in a list of 35 apps.
Out of them, we left out another three apps because they were: (a1) crashing \droidperm with \verb|OutOfMemoryError|, (a2) we couldn't build Android Studio project for it, required for manual inspection, (a3) was a duplciate of another app in the corpus.
Our final corpus, used for all three RQs, consists of the remaining 32 apps.
It includes 7 permissions and 41 (app, permission) instances.
The average execution time per app was 3.6 minutes, on a quad-core 3.5 Ghz processor with 16 GB of RAM.

\subsection{RQ1: What is the recall of \droidperm?}

We used a combination of helper tools and manual inspection to decide which of the undetected sensitives are truly unreachable and which are false negatives.
This hybrid approach allowed us to decrease the amount of manual work and reduce the chance of human errors.

\MyPara{Helper tools}
We already have the Permission Collector tool (\S\ref{sec:coverage-evaluation}) that finds all sensitives in the app.
The difference between all sensitives and those detected by \droidperm would make potentially undetected sensitives.
They need further inspection to be categorized as either false negatives or true negatives.
To filter out some of them automatically, we designed a simple class hierarchy-based reachability analysis tool that starts from the same dummy main method as \droidperm and uses class hierarchy data alone to resolve virtual calls.
These tools together divide the set of all sensitives into three partitions: (a) unreachable, (b) CHA-reachable but undetected and (c) detected by \droidperm.

\MyPara{Manual inspection}
We next manually inspected all CHA-reachable but undetected sensitive instances in our corpus, to determine which of them represent actually reachable sensitives, and thus false negatives.
For field sensitives we first inspected whether they are actually used to access a protected resource.
During this process, we occasionally found new methods that accept sensitive fields to access protected resources.
We used them to extend our specification of parametric sensitives.
Then, for both field and method sensitives we searched for possible call paths from callbacks.
If there was at least one feasible path, we considered the sensitive a false negative.
If we could not determine if a sensitive is reachable due to code complexity, we gave it benefit of the doubt and counted as reachable.

\begin{table}
\begin{tabularx}{\linewidth}{| X | c| c | c |}
Sensitives        & Method & Field & Total \\
\hline
Detected 											& 49 & 23 &	72 \\
Undetected, valid sensitive 	& 8 & 1 &	9 \\
Undetected, invalid sensitive & 0 & 6 &	6 \\
CHA-unreachable 									& 9 & 0 &	9 \\

\hline
Recall 	& 86\% & 96\% &	89\% \\

\end{tabularx}
\caption{Summary for recall evaluation}
\label{t:eval-recall}
\end{table}

\MyPara{Results}
Table~\ref{t:eval-recall} presents the results. We collected 72 detected sensitives, 9 valid but undetected sensitives and 6 invalid sensitives.
Since it is hard to count the missing paths for valid undetected sensitives, we consider the unit of our evaluation a sensitive.
This leads to a recall of $\frac{72}{72+9} \times 100\% = 89\% $.
Field sensitives that were filtered out all represent references to fields that do not result in access to protected resources.
Valid undetected methods are caused by one of the two reasons: (a) call graph incompleteness due to lack of access to android internals (\S\ref{sec:cg-augmentation}) or (b) benefit of the doubt.
The only instance of a valid undetected field is caused by a more complex type of parametric sensitive that depends of two parameters, that \droidperm does not support.
Adding support would be a localized engineering task.

\subsection{RQ2: What is the precision of \droidperm?}

In this section we evaluate the precision of \droidperm, as a ratio of sensitives for which all detected paths are valid divided by the total number of detected sensitives.
We take the sensitive as unit of precision, rather than path or permission request insertion location, because the total number of paths and reported request locations differs widely from one app to another. For most apps \droidperm reports only one path, while for a few - 20-30 paths. Thus, we prevent our evaluation results to be dominated by a few apps.

Here we also used a mix of tools and manual inspection.

\MyPara{Helper tools}
For every path from callback to sensitive \droidperm computes one more boolean flag: whether the path is ambiguous.
A path is ambiguous if it contains at least one method call that in the path's context can be resolved to more than one edge.
We considered non-ambiguous paths valid, and manually inspected ambiguous paths only.
This dramatically reduced the amount of required effort.
Out of 72 detected sensitives only 31 contained ambiguous paths, and for these 31 most of the paths were non-ambiguous.

\MyPara{Manual inspection}
For each ambiguous path we first identified ambiguous method calls.
For each of them, we performed manual dataflow analysis of the call target, to determine possible types (e.g. real points-to information).
For this we used the feature "Analyze Dataflow to Here" from Android Studio.
If manually inferred points-to set was not compatible with the respective edge in the analyzed path, we would qualify the path as invalid.

\begin{table}
\begin{tabularx}{\linewidth}{| X | c| c | c |}
Sensitives        & Method & Field & Total \\
\hline
Detected 											& 49 & 23 &	72 \\
With non-ambiguous paths only	& 28 & 13 & 41 \\
With non-ambiguous and valid ambiguous paths & 19 & 9 & 28\\
With invalid paths & 2 & 1 & 3 \\
Total with valid paths only & 47 & 22 & 69 \\
\hline
Precision 	& 96\% & 96\% &	96\% \\

\end{tabularx}
\caption{Summary for precision evaluation}
\label{t:eval-precision}
\end{table}

\MyPara{Results}
Table \ref{t:eval-precision} summarizes the results.
We found only 3 sensitives with invalid paths out of the total of 72, leading to a precision of $\frac{72-3}{72} \times 100\% = 96\% $.
All invalid paths were caused by insufficient context sensitivity.

\subsection{RQ3: What is the effect of 1-CFA points-to refinement on precision?}

\MyPara{Methodology}
To answer this question we ran \droidperm with GEOM points-to refinement disabled and compared the results with those of fully enabled \droidperm. We'll refer to the two configurations as 0-CFA and 1-CFA.

\MyPara{Results}
Configuration 0-CFA reported additional paths in 3 apps for 11 sensitives.
Of them, 2 sensitives in one app already had invalid paths with 1-CFA, from RQ2.
Thus, the total number of sensitives with invalid paths was $11-2+3 = 12$ .
Since 1-CFA is expected to be decidedly more precise than 0-CFA, we consider all the extra paths to be invalid. The number of additional paths per sensitive was highly nonuniform, ranging from 1 to 170.
The precision for 0-CFA configuration, with unit of precision being a sensitive, is $\frac{72-12}{72} \times 100\% = 83\% $
This proves that 1-CFA points-to analysis reduces the number of improperly detected sensitives 4-fold.

Detailed evaluation results can be found on: \cite{studyURL}.


\section{Studies derived from \dpspec}\label{sec:studies-from-dpspec}

In this section we present two studies derived from \dpspec and associated tools.
In \S\ref{sec:axplorer} we perform a detailed comparison between \dpspec and the current state of the art --- Axplorer~\cite{axplorer}.
In \S\ref{sec:over-privileged-apps} we present a small scale study of declaratively over-privileged apps.

\subsection{Comparison with Axplorer}
\label{sec:axplorer}
Next we compare \dpspec with the most recent work in this area: Axplorer~\cite{axplorer}.
For a fair comparison we had to bring them to a common denominator.
Axplorer specification contains 307 mappings for methods with both sensitive and non-sensitive permissions.
Since we only study sensitive permissions, we filtered the specification to exclude the rest, thus only leaving 45 mappings.
In addition, Axplorer does not support field sensitives.
Thus, we excluded field sensitives from \droidperm specification, leaving the 120 method sensitives only.
The two specifications have 27 common mappings, with no discrepancies w.r.t. referred permissions.

The remaining 18 mappings from Axplorer were new for \dpspec.
We analyzed all of them to determine whether they represent true sensitives.
For this we used the following working definition of a sensitive, a slight variation of that from \S\ref{sec:spec-javadoc} : A sensitive is a method that either (a) throws \texttt{SecurityException} when called without appropriate permissions or
(b) could be used to access known protected resources, without having to access other sensitives first.
According to this definition, only 7 of 18 extra mappings in Axplorer are undoubtedly sensitives: one because it is documented to throw \texttt{SecurityException}, and 6 by virtue of their functionality. Out of them three are deprecated.
Another one is modeled by \droidperm in a more fine-grained fashion, as a parametric sensitive.
The remaining 10 don't mention criteria (a) in their documentation and their functionality doesn't seem to satisfy criteria (b).

For example, \texttt{LocationManager.removeUpdates()} is explicitly documented to require permissions only up to API22, thus it is not a sensitive on API23.
Other methods represent either similar cleanup operations, status check operations like \texttt{SipManager.isOpened()}
or other operations that don't grant access to protected resources.
Yet all these methods plausibly use permission checks under the hood.
After reflecting on these findings, we decided not to merge Axplorer specification with ours.

\MyPara{Coverage comparison}
A larger permission specification does not necessarily mean better coverage.
Thus, we also compared the permissions coverage that the two specifications produce on our app corpus.
We excluded from both the method that is modeled differently.
In addition, we excluded from \dpspec mappings produced by manual inspection (\S\ref{sec:spec-apps}) and all mappings for fields.
Axplorer specification leads to 32 MCS permission instances, covering only 3 permissions.
In contrast, \droidperm produced 69 MCS instances covering 7 permissions.
Thus, even for sensitive methods alone it has more than twice the coverage of Axplorer.

\subsection{Declaratively over-privileged apps}
\label{sec:over-privileged-apps}
We found one more interesting pattern in Permission Collector data.
Declaratively over-privileged apps are those that declare permissions in the manifest but don't use them.
These permissions which are neither referred in the code nor consumed by sensitives.
Such app will not be over-privileged on Android 6, but will be on Android 5 or earlier.

Permission Collector found 24 such apps.
In 14 of them, over-declared permissions are in the same groups as other, properly used permissions.
For example, when the app properly uses \texttt{ACCESS\_COARSE\_LOCATION}, but also declares \texttt{ACCESS\_FINE\_LOCATION}.
In such cases extra permissions don't create any visible effects for the user
This is because Android only displays permission groups to the user, but not individual permissions.

In 10 other cases, over-declared permissions were from separate groups, making them visible to the user.
Such permissions could make the user wonder why they are needed and be skeptical to install the app.
We submitted issues for all apps in this group that have an issue tracker on github, 6 in total.
In 3 cases developers replied and fixed the issue.
None of the issues were rejected.
Developers gave various explanations for where these permissions come from.
In some cases they initially considered using the permission, added it to the manifest, but afterwards found it unnecessary.
In one case the permission was introduced in the manifest by Android build system, without any intervention from developer, simply because app linked to certain libraries \cite{web-perm-overprivileged}.
Thus, even careful developers might end up building over-privileged apps without being aware.

This study confirms that declaratively over-privileged apps, widely studied before\cite{perm-demystified, perm-overprivilege-2, perm-overprivilege-3}, remain frequent on Android 6.
Even though developers had a chance to review what permissions apps really need when performing the migration.
Over-declared permissions are only an issue when either the phone or the app run or is built for older Android versions.
Thus, there is a necessity to migrate both phones and apps to newer OS versions, and to provide tools that facilitate apps migration.

There is one more application of this section' study.
Android Studio has a code inspection/warning called \textit{"Calling a method that requires a permission without having declared that permission in the manifest"}.
The inspection uses permission annotations (\S\ref{sec:spec-anno}), which we showed to be vastly incomplete.
Thus, it is of limited usability.
Using the extensive specification from this study would allow both a wider applicability of the inspection but also the opposite inspection.
Now Android Studio could issue a warning if permission is declared without being used, with small false positive rates.
We provided \dpspec to Google engineers, to help them update permission annotations.


\section{Related Work}\label{sec:related_work}

We divide the related work in two categories: research dedicated to inferring permission mappings for Android, and research related to runtime permission usages in mobile apps.

\subsection{Inferring Permission Mappings}

Stowaway~\cite{perm-demystified} was the first research aimed to produce a comprehensive permission specification.
They used automatic test generation to exercise Android API, and modified the internal permission check mechanism to log all permission checks within Android.
The resulting specification still contained a number of inconsistencies which required manual inspection.
One inconsistency is methods whose permissions depend on received argument, which we model as parametric sensitives.
The final specification was precise yet incomplete.
In addition, Stowaway only analyzed Android 2.2.

The next major effort in this area was PScout~\cite{pscout}. Similarly to \droidperm it constructs a call graph and performs reachability traversal.
Unlike \droidperm it analyzes Android Framework code and looks for Android internal permission checks.
Due do difficulty to scale analysis to the whole Android, PScout used the imprecise CHA-based call graph generation.
PScout produces a vastly larger specification than Stowaway, yet it suffers from imprecision.
The evaluation they perform is coarse-grained.
For a corpus of apps, permissions declared in the manifest are compared with permissions inferred from PScout specification. The paper states the upper bound of false mappings of 7\%.
It is easy to see what could be missed by this evaluation.
If two methods are always used together, yet only one of them requires permissions, the evaluation won't be able to determine which one is that.

At the moment \droidperm project started, PScout was the most recent permission specification available.
We considered to use it, but soon realized the precision problem.
A quick look at PScout specification for API22\footnote{\url{https://github.com/zyrikby/PScout/blob/master/results/API\_22/publishedapimapping}}
reveals many implausible mappings, like \verb|android.app.UiAutomation: getWindows()| requiring \verb|CHANGE_WIFI_STATE|.
Thus we decided to build our own specification.

The state of the art in generating an automatic permission specification is Axplorer~\cite{axplorer}.
It follows the same key idea as PScout.
Axplorer builds on top of PScout through more precise modeling of Android internals.
Also, instead of CHA, they use the more precise points-to based call graph.
To overcome scalability problem, Axplorer is run on more powerful hardware.
Being the most recent, this is the only project to analyze Android 6.
Axplorer was evaluated relatively to PScout only.
The evaluation revealed superior precision, both due to correct engineering and better analysis.
As we show in this paper, Axplorer specification is fairly precise, yet contains few mappings for dangerous permissions, ultimately producing less than half the coverage of permissions used by apps compared to \dpspec.

Copes~\cite{copes} is a tool that discovers over-privileged apps for Android 2.2.
This work also produces a permission specification using reachability analysis of Android Framework, similar to PScout.
They use SPARK to generate the call graph, but have a simplified model of Android internals.
Copes specification was not evaluated.
To detect over-privilege, Copes project includes a tool similar to our Permission Collector (\S\ref{sec:coverage-evaluation}).

In contrast with previous efforts, \dpspec is built from available documentation sources, plus to a small degree from permission usages in apps.
Its precision is bounded by the correction of this documentation and our methodology.
For \droidperm, we use a precise evaluation of each individual path, capable of producing absolute precision and recall values.
In principle, similar evaluation could be used for other reachability analyses, including PScout and Axplorer.

In a slightly different context, Kratos~\cite{kratos} is another tool that performs reachability analysis on Android Framework.
Kratos searches inconsistencies in permission check mechanism that could allow apps to access resources without appropriate permissions.

\subsection{Runtime Permissions in Mobile Apps}

The only other tool to our knowledge that analyzes runtime permissions for Android is revDroid~\cite{revDroid}.
It searches for bugs that arise when permissions are checked too far away from where they are used.
These bugs can lead to crashes when permissions are revoked while app is running.
Yet revDroid incorrectly assumes that the bug happens any time a the permission check and the sensitive are in different callbacks.
Similar to \droidperm, revDroid is based on Soot and FlowDroid platform.
Different from \droidperm, revDroid uses PScout specification, context-insensitive SPARK call graph and default FlowDroid entry point generation.
The paper doesn't present any solution for call graph incompleteness.
As a result of these shortcomings, the paper finds that 46\% of sensitives have their permissions checked improperly, which we believe is a vast overstatement.
revDroid paper has no evaluation.

We are also interested in extending \droidperm with analysis of permission checks. This requires significantly more complex modeling of Android components lifecycle for good precision.

Livshits and Jung~\cite{windows-phone-perm} proposed a tool for instrumenting Windows Phone (WP) apps with permission prompts.
There is one conceptual difference between this tool and \droidperm.
On WP permission prompts can be placed anywhere in the app binary code. They cannot be placed only inside dll libraries.
In contrast, on Android requests have to be propagated to the calling callback.
As a result, WP tool places prompts straight in front of the protected resource in most cases.
Being less concerned with precision of interprocedural propagation, they use CHA call graph, which also does not suffer from incompleteness.

\section{Conclusions}\label{sec:conclusion}

We first presented \dpspec, produced from various documentation sources using straightforward tools and careful interpretation on ambiguities.
It is essentially the first attempt to gather in one piece all the documentation on permissions available in Android SDK.
Yet, even the simple evaluation and comparison that we performed led to a number of conclusions, somewhat expected in outcome but surprising in magnitude.
First, we have a definitive evidence that Android SDK documentation is incomplete and this complicates the life of developers when dealing with permissions.
Second, even the state of the art tools for building permission specifications from source code produce vastly incomplete results, even when carefully modeling android internals.

\dpspec complements these efforts by providing a better benchmark with which to compare, so that limitations could be easily spotted and discrepancies could be explained, thus driving future advancements in the area.
In addition, it is ready to be used in any application in which a permission specification is needed.
We also provided the specification to Google engineers to help them extend permission annotations.
With relatively little effort, it is possible to maintain up to date specifications for newer versions of Android.

We presented \droidperm, a highly complete and precise analysis for recommending permission request locations.
In fact Android Studio already has an inspection with the same goal, but it is severely limited in at least two ways.
First it only supports sensitives in annotations. Second, the inspection reports issues at the location of sensitive, without propagating it inter-procedurally.
\dpspec and \droidperm are out of the box solutions ready to be plugged into this inspection for dramatically better results.

We envision \droidperm as the first step towards a comprehensive tool suite that would allow developers (a) automatically migrate all protected resources to runtime model, (b) maintain correct permission requests for evolving apps that have already been migrated.
For this, we intend to extend \droidperm with (i) storage permissions, (ii) an inter-component communication model (for example by integration with IccTA~\cite{iccta}), and (iii) an analysis of permission checks.
The extended tool should be able to correctly detect bugs studied in~\cite{revDroid} as well as a number of other issues related to runtime permissions, thus providing long-term benefits for Android community.

\newpage
\bibliographystyle{IEEEtran}
\bibliography{sections/references}

\begin{thebibliography}{10}
\providecommand{\url}[1]{#1}
\csname url@samestyle\endcsname
\providecommand{\newblock}{\relax}
\providecommand{\bibinfo}[2]{#2}
\providecommand{\BIBentrySTDinterwordspacing}{\spaceskip=0pt\relax}
\providecommand{\BIBentryALTinterwordstretchfactor}{4}
\providecommand{\BIBentryALTinterwordspacing}{\spaceskip=\fontdimen2\font plus
\BIBentryALTinterwordstretchfactor\fontdimen3\font minus
  \fontdimen4\font\relax}
\providecommand{\BIBforeignlanguage}[2]{{%
\expandafter\ifx\csname l@#1\endcsname\relax
\typeout{** WARNING: IEEEtran.bst: No hyphenation pattern has been}%
\typeout{** loaded for the language `#1'. Using the pattern for}%
\typeout{** the default language instead.}%
\else
\language=\csname l@#1\endcsname
\fi
#2}}
\providecommand{\BIBdecl}{\relax}
\BIBdecl

\bibitem{perm-msr-13}
\BIBentryALTinterwordspacing
R.~Stevens, J.~Ganz, V.~Filkov, P.~Devanbu, and H.~Chen, ``Asking for (and
  about) permissions used by android apps,'' in \emph{Proceedings of the 10th
  Working Conference on Mining Software Repositories}, ser. MSR '13.\hskip 1em
  plus 0.5em minus 0.4em\relax Piscataway, NJ, USA: IEEE Press, 2013, pp.
  31--40. [Online]. Available:
  \url{http://dl.acm.org/citation.cfm?id=2487085.2487093}
\BIBentrySTDinterwordspacing

\bibitem{perm-demystified}
\BIBentryALTinterwordspacing
A.~P. Felt, E.~Chin, S.~Hanna, D.~Song, and D.~Wagner, ``Android permissions
  demystified,'' in \emph{Proceedings of the 18th ACM Conference on Computer
  and Communications Security}, ser. CCS '11.\hskip 1em plus 0.5em minus
  0.4em\relax New York, NY, USA: ACM, 2011, pp. 627--638. [Online]. Available:
  \url{http://doi.acm.org/10.1145/2046707.2046779}
\BIBentrySTDinterwordspacing

\bibitem{perm-risks-benefits}
\BIBentryALTinterwordspacing
B.~P. Sarma, N.~Li, C.~Gates, R.~Potharaju, C.~Nita-Rotaru, and I.~Molloy,
  ``Android permissions: A perspective combining risks and benefits,'' in
  \emph{Proceedings of the 17th ACM Symposium on Access Control Models and
  Technologies}, ser. SACMAT '12.\hskip 1em plus 0.5em minus 0.4em\relax New
  York, NY, USA: ACM, 2012, pp. 13--22. [Online]. Available:
  \url{http://doi.acm.org/10.1145/2295136.2295141}
\BIBentrySTDinterwordspacing

\bibitem{perm-sleeping-android}
\BIBentryALTinterwordspacing
J.~Sellwood and J.~Crampton, ``Sleeping android: The danger of dormant
  permissions,'' in \emph{Proceedings of the Third ACM Workshop on Security and
  Privacy in Smartphones \&\#38; Mobile Devices}, ser. SPSM '13.\hskip 1em plus
  0.5em minus 0.4em\relax New York, NY, USA: ACM, 2013, pp. 55--66. [Online].
  Available: \url{http://doi.acm.org/10.1145/2516760.2516774}
\BIBentrySTDinterwordspacing

\bibitem{perm-remystified}
\BIBentryALTinterwordspacing
P.~Wijesekera, A.~Baokar, A.~Hosseini, S.~Egelman, D.~Wagner, and K.~Beznosov,
  ``Android permissions remystified: A field study on contextual integrity,''
  in \emph{Proceedings of the 24th USENIX Conference on Security Symposium},
  ser. SEC'15.\hskip 1em plus 0.5em minus 0.4em\relax Berkeley, CA, USA: USENIX
  Association, 2015, pp. 499--514. [Online]. Available:
  \url{http://dl.acm.org/citation.cfm?id=2831143.2831175}
\BIBentrySTDinterwordspacing

\bibitem{so-sensitive-list}
\BIBentryALTinterwordspacing
(2015) Mapping between android permissions to corresponding api calls methods.
  [Online]. Available: \url{http://stackoverflow.com/a/24019120}
\BIBentrySTDinterwordspacing

\bibitem{so-sensitive-list-2}
\BIBentryALTinterwordspacing
(2016) How to find required android marshmallow runtime permissions in code?
  [Online]. Available: \url{http://stackoverflow.com/q/34679400}
\BIBentrySTDinterwordspacing

\bibitem{so-sensitive-list-3}
\BIBentryALTinterwordspacing
(2015) Migrating to runtime permissions: How do you find all the current
  permission uses? [Online]. Available:
  \url{http://stackoverflow.com/q/32656505}
\BIBentrySTDinterwordspacing

\bibitem{pscout}
\BIBentryALTinterwordspacing
K.~W.~Y. Au, Y.~F. Zhou, Z.~Huang, and D.~Lie, ``Pscout: Analyzing the android
  permission specification,'' in \emph{Proceedings of the 2012 ACM Conference
  on Computer and Communications Security}, ser. CCS '12.\hskip 1em plus 0.5em
  minus 0.4em\relax New York, NY, USA: ACM, 2012, pp. 217--228. [Online].
  Available: \url{http://doi.acm.org/10.1145/2382196.2382222}
\BIBentrySTDinterwordspacing

\bibitem{axplorer}
\BIBentryALTinterwordspacing
M.~Backes, S.~Bugiel, E.~Derr, P.~McDaniel, D.~Octeau, and S.~Weisgerber, ``On
  demystifying the android application framework: Re-visiting android
  permission specification analysis,'' in \emph{25th USENIX Security Symposium
  (USENIX Security 16)}.\hskip 1em plus 0.5em minus 0.4em\relax Austin, TX:
  USENIX Association, 2016, pp. 1101--1118. [Online]. Available:
  \url{https://www.usenix.org/conference/usenixsecurity16/technical-sessions/presentation/backes_android}
\BIBentrySTDinterwordspacing

\bibitem{url-normal-and-dangerous-perm}
\BIBentryALTinterwordspacing
(2015) Normal and dangerous permissions. [Online]. Available:
  \url{https://developer.android.com/guide/topics/permissions/requesting.html}
\BIBentrySTDinterwordspacing

\bibitem{url-perm-best-practices}
\BIBentryALTinterwordspacing
(2015) Normal and dangerous permissions. [Online]. Available:
  \url{https://developer.android.com/training/articles/user-data-permissions.html}
\BIBentrySTDinterwordspacing

\bibitem{soot}
\BIBentryALTinterwordspacing
R.~Vall{\'e}e-Rai, P.~Co, E.~Gagnon, L.~Hendren, P.~Lam, and V.~Sundaresan,
  ``Soot: A java bytecode optimization framework,'' in \emph{CASCON First
  Decade High Impact Papers}, ser. CASCON '10.\hskip 1em plus 0.5em minus
  0.4em\relax Riverton, NJ, USA: IBM Corp., 2010, pp. 214--224. [Online].
  Available: \url{http://dx.doi.org/10.1145/1925805.1925818}
\BIBentrySTDinterwordspacing

\bibitem{flowdroid}
\BIBentryALTinterwordspacing
S.~Arzt, S.~Rasthofer, C.~Fritz, E.~Bodden, A.~Bartel, J.~Klein, Y.~Le~Traon,
  D.~Octeau, and P.~McDaniel, ``Flowdroid: Precise context, flow, field,
  object-sensitive and lifecycle-aware taint analysis for android apps,'' in
  \emph{Proceedings of the 35th ACM SIGPLAN Conference on Programming Language
  Design and Implementation}, ser. PLDI '14.\hskip 1em plus 0.5em minus
  0.4em\relax New York, NY, USA: ACM, 2014, pp. 259--269. [Online]. Available:
  \url{http://doi.acm.org/10.1145/2594291.2594299}
\BIBentrySTDinterwordspacing

\bibitem{paddle}
O.~Lhotak, ``Program analysis using binary decision diagrams,'' Ph.D.
  dissertation, Montreal, Que., Canada, Canada, 2006, aAINR25195.

\bibitem{SPARK}
\BIBentryALTinterwordspacing
O.~Lhot\'{a}k and L.~Hendren, ``Scaling java points-to analysis using spark,''
  in \emph{Proceedings of the 12th International Conference on Compiler
  Construction}, ser. CC'03.\hskip 1em plus 0.5em minus 0.4em\relax Berlin,
  Heidelberg: Springer-Verlag, 2003, pp. 153--169. [Online]. Available:
  \url{http://dl.acm.org/citation.cfm?id=1765931.1765948}
\BIBentrySTDinterwordspacing

\bibitem{geom}
\BIBentryALTinterwordspacing
X.~Xiao and C.~Zhang, ``Geometric encoding: Forging the high performance
  context sensitive points-to analysis for java,'' in \emph{Proceedings of the
  2011 International Symposium on Software Testing and Analysis}, ser. ISSTA
  '11.\hskip 1em plus 0.5em minus 0.4em\relax New York, NY, USA: ACM, 2011, pp.
  188--198. [Online]. Available:
  \url{http://doi.acm.org/10.1145/2001420.2001443}
\BIBentrySTDinterwordspacing

\bibitem{revDroid}
\BIBentryALTinterwordspacing
Z.~Fang, W.~Han, D.~Li, Z.~Guo, D.~Guo, X.~S. Wang, Z.~Qian, and H.~Chen,
  ``revdroid: Code analysis of the side effects after dynamic permission
  revocation of android apps,'' in \emph{Proceedings of the 11th ACM on Asia
  Conference on Computer and Communications Security}, ser. ASIA CCS '16.\hskip
  1em plus 0.5em minus 0.4em\relax New York, NY, USA: ACM, 2016, pp. 747--758.
  [Online]. Available: \url{http://doi.acm.org/10.1145/2897845.2897914}
\BIBentrySTDinterwordspacing

\bibitem{studyURL}
\BIBentryALTinterwordspacing
(2017, Feb) Droidperm study. [Online]. Available: \url{anonymous}
\BIBentrySTDinterwordspacing

\bibitem{web-perm-overprivileged}
\BIBentryALTinterwordspacing
(2015, Jun) Hey, where did these permissions come from? [Online]. Available:
  \url{https://commonsware.com/blog/2015/06/25/hey-where-did-these-permissions-come-from.html}
\BIBentrySTDinterwordspacing

\bibitem{perm-overprivilege-2}
\BIBentryALTinterwordspacing
A.~Atzeni, T.~Su, M.~Baltatu, R.~D'Alessandro, and G.~Pessiva, ``How dangerous
  is your android app?: An evaluation methodology,'' in \emph{Proceedings of
  the 11th International Conference on Mobile and Ubiquitous Systems:
  Computing, Networking and Services}, ser. MOBIQUITOUS '14.\hskip 1em plus
  0.5em minus 0.4em\relax ICST, Brussels, Belgium, Belgium: ICST (Institute for
  Computer Sciences, Social-Informatics and Telecommunications Engineering),
  2014, pp. 130--139. [Online]. Available:
  \url{http://dx.doi.org/10.4108/icst.mobiquitous.2014.257832}
\BIBentrySTDinterwordspacing

\bibitem{perm-overprivilege-3}
\BIBentryALTinterwordspacing
J.~Sellwood and J.~Crampton, ``Sleeping android: The danger of dormant
  permissions,'' in \emph{Proceedings of the Third ACM Workshop on Security and
  Privacy in Smartphones \&\#38; Mobile Devices}, ser. SPSM '13.\hskip 1em plus
  0.5em minus 0.4em\relax New York, NY, USA: ACM, 2013, pp. 55--66. [Online].
  Available: \url{http://doi.acm.org/10.1145/2516760.2516774}
\BIBentrySTDinterwordspacing

\bibitem{copes}
\BIBentryALTinterwordspacing
A.~Bartel, J.~Klein, Y.~Le~Traon, and M.~Monperrus, ``Automatically securing
  permission-based software by reducing the attack surface: An application to
  android,'' in \emph{Proceedings of the 27th IEEE/ACM International Conference
  on Automated Software Engineering}, ser. ASE 2012.\hskip 1em plus 0.5em minus
  0.4em\relax New York, NY, USA: ACM, 2012, pp. 274--277. [Online]. Available:
  \url{http://doi.acm.org/10.1145/2351676.2351722}
\BIBentrySTDinterwordspacing

\bibitem{kratos}
Y.~Shao, Q.~A. Chen, Z.~M. Mao, J.~Ott, and Z.~Qian, ``Kratos: Discovering
  inconsistent security policy enforcement in the android framework,'' in
  \emph{NDSS}, 2016.

\bibitem{windows-phone-perm}
\BIBentryALTinterwordspacing
B.~Livshits and J.~Jung, ``Automatic mediation of privacy-sensitive resource
  access in smartphone applications,'' in \emph{Proceedings of the 22Nd USENIX
  Conference on Security}, ser. SEC'13.\hskip 1em plus 0.5em minus 0.4em\relax
  Berkeley, CA, USA: USENIX Association, 2013, pp. 113--130. [Online].
  Available: \url{http://dl.acm.org/citation.cfm?id=2534766.2534777}
\BIBentrySTDinterwordspacing

\bibitem{iccta}
\BIBentryALTinterwordspacing
L.~Li, A.~Bartel, T.~F. Bissyand{\'e}, J.~Klein, Y.~Le~Traon, S.~Arzt,
  S.~Rasthofer, E.~Bodden, D.~Octeau, and P.~McDaniel, ``Iccta: Detecting
  inter-component privacy leaks in android apps,'' in \emph{Proceedings of the
  37th International Conference on Software Engineering - Volume 1}, ser. ICSE
  '15.\hskip 1em plus 0.5em minus 0.4em\relax Piscataway, NJ, USA: IEEE Press,
  2015, pp. 280--291. [Online]. Available:
  \url{http://dl.acm.org/citation.cfm?id=2818754.2818791}
\BIBentrySTDinterwordspacing

\end{thebibliography}

\end{document}